# Planckian Interacting Massive Particles as Dark Matter

Mathias Garny,[1,*] McCullen Sandora,[2,†] and Martin S. Sloth[2,‡]
[1]*CERN Theory Division, CH-1211 Geneva 23, Switzerland*
[2]*CP3-Origins, Center for Cosmology and Particle Physics Phenomenology,
University of Southern Denmark, Campusvej 55, 5230 Odense M, Denmark*
(Received 20 November 2015; published 10 March 2016)

The standard model could be self-consistent up to the Planck scale according to the present measurements of the Higgs boson mass and top quark Yukawa coupling. It is therefore possible that new physics is only coupled to the standard model through Planck suppressed higher dimensional operators. In this case the weakly interacting massive particle miracle is a mirage, and instead minimality as dictated by Occam's razor would indicate that dark matter is related to the Planck scale, where quantum gravity is anyway expected to manifest itself. Assuming within this framework that dark matter is a Planckian interacting massive particle, we show that the most natural mass larger than $0.01 M_p$ is already ruled out by the absence of tensor modes in the cosmic microwave background (CMB). This also indicates that we expect tensor modes in the CMB to be observed soon for this type of minimal dark matter model. Finally, we touch upon the Kaluza-Klein graviton mode as a possible realization of this scenario within UV complete models, as well as further potential signatures and peculiar properties of this type of dark matter candidate. This paradigm therefore leads to a subtle connection between quantum gravity, the physics of primordial inflation, and the nature of dark matter.

DOI: 10.1103/PhysRevLett.116.101302

*Introduction.*—The recent measurements of the Higgs boson mass and top quark Yukawa coupling, together with an absence of new physics beyond the standard model (SM) in the experiments at LHC, leave open the possibility that new physics will not have to manifest itself below the Planck scale [1]. Within the traditional mindset of naturalness in the 't Hooft sense [2], this would be problematic and leave open the hierarchy problem, with no reasonable explanation of why the electroweak scale is so much smaller than the Planck scale. However, with the advent of the string landscape, an alternative approach to naturalness has been developed. In this framework, the constants of nature which are required to be unnaturally small in order for complex life to emerge, could simply be a consequence of environmental selection [3]. For example, it has been argued that the properties of nuclei and atoms would not allow for a complex chemistry to exist if the electroweak scale is too far away from the confinement scale of QCD [4–6].

If indeed the smallness of the electroweak scale is a consequence of environmental selection and the hierarchy problem is resolved by anthropic arguments, then there is no compelling reason to believe that dark matter (DM) should be related to the electroweak scale, and we might be forced to give up the weakly interacting massive particle (WIMP) [7] paradigm for DM. In this case, if we do not want to introduce any new energy scale specifically for explaining DM, the only required scale for new physics to appear is the Planck scale $M_p = 1.2 \times 10^{19}$ GeV, or perhaps the grand unified theory (GUT) scale, $10^{-3} M_p$. We are therefore led to the possibility that in a minimal framework DM must be related to the Planck scale or GUT scale, and DM interactions with the SM are controlled by Planck suppressed higher dimensional operators. Thus, within this alternative approach to naturalness, a possible "natural" DM candidate is a Planckian interacting dark matter particle (PIDM).

While we assume that, within the minimal PIDM paradigm, DM has gravitational interactions only and a mass not too far below the Planck or GUT scales, it is certainly possible that the hidden sector is more complicated. For example, if there are additional symmetries that protect the PIDM mass, it could be much lighter than the Planck or GUT scales. However, taking Occam's razor as a principle, in this Letter we study the simplest case from a low energy field theory point of view, where the hidden sector is minimal, and the PIDM mass is close to the Planck scale—we call this the minimal PIDM paradigm, as opposed to the nonminimal PIDM paradigm where the mass can be lighter.

This type of scenario has not received much attention previously in the literature, possibly because it has been assumed to be very hard to test. There certainly will not be such dramatic signatures as in direct DM detection experiments, and the chance of seeing a decaying PIDM in an indirect detection experiment is very slim. However, an interesting consequence of the minimal PIDM paradigm is that in order to produce enough of such a heavy, very weakly coupled particle, the reheating temperature of inflation has to be relatively high. This implies a detectable primordial tensor-to-scalar ratio $r$ in the cosmic microwave background (CMB) power spectrum. Below, we show that the current upper bound on $r$ from the Planck satellite and the Keck array [8,9] already constrains the PIDM mass to be below $0.01 M_p$. In the minimal PIDM scenario, we therefore expect tensor modes to be discovered soon, or else we will be under pressure to give up the minimal PIDM paradigm.

Some comments relating this model to other weakly coupled dark matter candidates in the literature follow. There is the WIMPZILLA [10], consisting of a heavy particle but with couplings that are not as suppressed, so that the cross section is set by the mass of the particle instead of the Planck mass. The SUPERWIMP [7], which is very weakly coupled but has a mass much smaller than the GUT scale, is produced by the decay of heavy particles that are themselves produced via freeze-out, like WIMPs. Also there is the feebly interacting massive particle (FIMP) [11], which is very weakly coupled, but in this instance produced by the freeze-in process. "Superheavy" dark matter with mass around $10^{13-15}$ GeV is produced due to the time-dependent background metric at the end of inflation, known as "gravitational production" [12–14]. The scenario we consider here could be viewed as a FIMPZILLA, with a very weakly coupled, very heavy particle being dominantly produced by freeze-in. Finally, we round up discussing the Kaluza-Klein graviton as a natural and minimal PIDM-like particle.

*Effective model.*—We assume that dark matter resides in a hidden sector that is connected only gravitationally to the SM. In the absence of selection effects, we should expect the dark matter mass to be natural, i.e., near the cutoff of the theory. To begin, we consider dark matter to be a single real scalar with mass $m_X$. Choosing the PIDM to be a fermion, or adding additional degrees of freedom, does not significantly affect our conclusions. The Lagrangian is

$$\mathcal{L} = \mathcal{L}_{\text{SM}} + \mathcal{L}_{\text{DM}} + \mathcal{L}_{\text{EH}} + \frac{1}{2} h^{\mu\nu}(T^{\text{SM}}_{\mu\nu} + T^{\text{DM}}_{\mu\nu}), \quad (1)$$

where $\mathcal{L}_{\text{SM}}$ denotes the SM Lagrangian, $\mathcal{L}_{\text{DM}}$ is the free Lagrangian of a massive scalar, and $\mathcal{L}_{\text{EH}}$ is the Einstein-Hilbert Lagrangian. The two sectors only communicate indirectly through gravity, which couples to their energy-momentum tensors. Since this coupling is uniquely determined from the equivalence principle, the mass of the PIDM is the only free parameter in this model. The $s$-channel graviton exchange diagram results in a $2 \to 2$ annihilation amplitude

$$\mathcal{M} = \frac{-i8\pi \langle p_1 | T^{\mu\nu}_{\text{SM}} | p_2 \rangle \langle k_a | (T^{\text{DM}}_{\mu\nu} - \frac{1}{2} g_{\mu\nu} T^{\text{DM}\alpha}_{\alpha}) | k_b \rangle}{M_p^2 (k_a + k_b)^2}, \quad (2)$$

where the on-shell energy momentum tensors are divergence free by diffeomorphism invariance, $(k_a + k_b)^\mu T_{\mu\nu} = 0$. Here, $M_p = 1.2 \times 10^{19}$ GeV is the nonreduced Planck mass.

*Production.*—Because of the very suppressed interactions, PIDMs are produced dominantly at the highest energy scales after inflation. We consider three contributions, corresponding to the period shortly after reheating, during reheating, and at the end of inflation. The first one can be treated similarly to a FIMP with nonrenormalizable interactions [11], and the last one corresponds to the gravitational production as in Refs. [12–14]. For the contribution during reheating, we use the description of the perturbative reheating dynamics in Ref. [15]

$$\frac{d\rho_\phi}{dt} = -3H(1+w)\rho_\phi - S,$$
$$\frac{d\rho_R}{dt} = -4H\rho_R + S + 2\langle \sigma v \rangle \langle E_X \rangle [n_X^2 - (n_X^{\text{eq}})^2],$$
$$\frac{dn_X}{dt} = -3H n_X - \langle \sigma v \rangle [n_X^2 - (n_X^{\text{eq}})^2] \quad (3)$$

governing the energy density of the inflaton (or, more generally, reheaton) $\rho_\phi$, radiation $\rho_R$, and number density of dark matter $n_X$. The dependence on the reheating dynamics enters via the term $S = \Gamma \rho_\phi$, with constant $\Gamma$, describing the inflaton decay into relativistic standard model degrees of freedom. Furthermore $\langle \sigma v \rangle$ is the usual thermally averaged $2 \to 2$ cross section for dark matter pair annihilation into SM particles. The inflaton field dynamics is parametrized by the effective equation of state $w$ (note that $w = 0$ in Ref. [15]). Instant preheating is reached for $\Gamma \to H_i$ (i.e., within one Hubble time) while $\Gamma \ll H_i$ for perturbative reheating, where $H_i$ is the Hubble rate at the initiation of reheating. The reheating temperature [defined by the condition $H(a) = \Gamma$] is

$$T_{\text{rh}} = \kappa_2 \gamma (M_p H_i)^{1/2}, \quad (4)$$

where $\kappa_2 = [45/(4\pi^3 g_{\text{rh}})]^{1/4} \approx 0.25$, $g_{\text{rh}}$ being the number of degrees of freedom at reheating, taken here to be that of the SM. This defines the constant $\gamma = \sqrt{\Gamma/H_i} \in (0,1)$, related to the efficiency of reheating. It can also be expressed in terms of the duration of reheating $N_{\text{rh}}$ as [16]

$$\gamma = \left(\frac{g_i}{g_{\text{rh}}}\right)^{1/4} e^{-\frac{3}{4}N_{\text{rh}}(1+w)}. \quad (5)$$

As we discuss below, the PIDM scenario is only viable for very efficient reheating, $\gamma \gtrsim 10^{-3}$. This can be achieved in several scenarios. Even with a strictly perturbative reheating, for which $\Gamma = g^2 m_\phi/(8\pi)$ (with coupling $g$ between the inflaton and the SM), we arrive at $\gamma \approx 0.2 g \sqrt{m_\phi/H_i}$. Thus, if the inflaton is not too weakly coupled to the radiation, and the mass of the inflaton during reheating is comparable to the Hubble rate, reheating can easily be efficient enough to produce the PIDM. Additionally, Ref. [17] laid out a nonperturbative scenario capable of achieving $\gamma \approx 1$, though this is somewhat nongeneric. While a quick reheating is necessary for our purposes, for perturbative reheating this generically exacerbates the naturalness problem in the inflaton sector [18]. This might motivate a generalization of the present study to include the more specific cases of nonperturbative reheating, but given that the inflaton sector already suffers a mild naturalness problem, we will here decouple the issue of naturalness in the inflaton sector from the present discussion and assume a generic perturbative reheating. The PIDM production during reheating will be subdominant within most of the viable mass range, so we do not expect our conclusions to depend strongly on these assumptions.

During reheating, the energy density in the inflaton field decays as $\rho_\phi \propto a^{-3(1+w)}$ until reheating stops and the Universe is dominated by a thermal plasma. Under these assumptions, the Hubble rate is

$$H \simeq H_i \begin{cases} (a/a_i)^{-3(1+w)/2}, & a < a_{\rm rh}, \\ \gamma^2 (a/a_{\rm rh})^{-2}, & a > a_{\rm rh}, \end{cases} \quad (6)$$

where $a_i = 1$ at the beginning of reheating, and $a_{\rm rh} = a_i \gamma^{-[4/3(1+\omega)]}$ is the scale factor when reheating is complete.

We are interested in relating $H_i$ to the Hubble rate measured from the CMB, corresponding to different stages of inflation. The inequality $H_{\rm CMB} > H_i$ is sufficient to place bounds on the PIDM scenario. The combined CMB bound on tensor modes, $r < 0.07$ (95% C.L.) [8,9], then translates into an upper bound on $H_i$

$$H_i < 6.6 \times 10^{-6} M_p \left(\frac{r}{0.1}\right)^{1/2}. \quad (7)$$

We also need the relation between temperature and the scale factor in order to calculate the abundance. During reheating the temperature behaves as [15]

$$T(a) = \frac{\kappa_1 (\gamma M_p H_i)^{1/2}}{(1 + 3/5w)^{1/4}} (a^{-3(1-w)/2} - a^{-4})^{1/4}, \quad (8)$$

where $\kappa_1 = [9/(2\pi^3 g_{\rm max})]^{1/4} \approx 0.20$, whereas after reheating

$$T(a) = T_{\rm rh} \frac{a_{\rm rh}}{a}. \quad (9)$$

Because of the Planck suppressed interactions, the PIDM abundance never dominates the energy density. Therefore, the evolution of the inflaton and radiation largely decouples from the Boltzmann equation governing the dark matter, which becomes

$$\frac{dX}{da} = \frac{a^2}{T_{\rm rh}^3 H(a)} \langle \sigma v \rangle (n_X^{\rm eq})^2, \quad (10)$$

for the dimensionless abundance $X = n_X a^3/T_{\rm rh}^3$. It holds when $n_X \ll n_{\rm eq}$, i.e., when the inverse annihilation process dominates. This turns out to be an excellent approximation during the period in which dark matter is produced, even when dark matter is not thermally distributed.

The thermally averaged cross section must be computed to continue. Using Eq. (2), and proceeding as in Ref. [19], we obtain

$$\langle \sigma v \rangle = N_0 \langle \sigma v \rangle_0 + N_{1/2} \langle \sigma v \rangle_{1/2} + N_1 \langle \sigma v \rangle_1,$$

$$\langle \sigma v \rangle_0 = \frac{\pi m_X^2}{M_p^4}\left[\frac{3}{5}\frac{K_1^2}{K_2^2} + \frac{2}{5} + \frac{4}{5}\frac{T}{m_X}\frac{K_1}{K_2} + \frac{8}{5}\frac{T^2}{m_X^2}\right],$$

$$\langle \sigma v \rangle_{1/2} = \langle \sigma v \rangle_1 = \frac{4\pi T^2}{M_p^4}\left\{\frac{2}{15}\left[\frac{m_X^2}{T^2}\left(\frac{K_1^2}{K_2^2} - 1\right)\right.\right.$$
$$\left.\left.+3\frac{m_X}{T}\frac{K_1}{K_2} + 6\right]\right\}, \quad (11)$$

where the subscripts denote the spin of the SM particle, and the $N_i$'s are the number of degrees of freedom of each type in the SM, namely, $N_0 = 4$, $N_{1/2} = 45$, and $N_1 = 12$. The $K_i$'s are modified Bessel functions, with argument $m_X/T(a)$, and the brackets asymptote to 1 for $m_X \gg T$, leaving the prefactor to display the nonrelativistic behavior ($s$ wave for scalars and $d$ wave otherwise). Since the matrix element squared scales like $|\mathcal{M}|^2 \sim E^4/M_p^4$, the phase space integration is dominated by the highest momenta that are not Boltzmann suppressed, $p \sim T$, such that corrections from quantum statistics are not large even for $m_X \ll T$.

Integrating Eq. (10) yields the final abundance $X_f$, assuming a vanishing initial abundance. We use Eqs. (4), (6), and (11), and send $a_f \to \infty$ since the production rate is exponentially suppressed by this point, making the integral insensitive to this region. We then look for combinations of parameters that yield the correct relic abundance through the expression

$$\Omega_X h^2 = Q \gamma^{4/(1+\omega)} \frac{m_X}{M_p} X_f,$$

$$Q = \frac{1}{8}\frac{T_{\rm rh}^3 M_p s_0}{s_{\rm rh} \rho_c} \approx 9.2 \times 10^{24}, \quad (12)$$

where $s_0(s_{\rm rh})$ is the entropy density today (at $T = T_{\rm rh}$) and $\rho_c = 1.88 \times 10^{-29}$ g/cm$^3$ is the critical density for $h = 1$. The factor $1/8$ is as introduced in Ref. [15], parametrizing the fact that entropy continues to be generated for a time after reheating. For the contribution from gravitational production at the end of inflation, which still needs to be added to Eq. (12), we use the result from Ref. [14]. In addition, the gravitational coupling to the inflaton sector can lead to another source of PIDMs, which is, however, model dependent: possible mechanisms are a perturbative decay for a light PIDM, and a contribution similar to Eq. (10) for a heavy PIDM. We checked that these contributions are subdominant within the accessible parameter space. We have also checked that the relevant temperatures fulfill $H < \alpha T$ within the viable range of parameters. Note that the naive estimate of the thermalization rate lies close to the value found in Ref. [20] for over-occupied initial conditions, which resemble most closely the spectrum expected from perturbative decay (cf. Refs. [21–23] for related discussions in the opposite regime). Nevertheless, a more detailed treatment of thermalization would be desirable for a precise determination of the relic density.

*Results and discussion.*—The level curves in Fig. 1 show the lines for which the PIDM abundance matches the cold dark matter density measured by Planck, $\Omega_c h^2 = 0.1198 \pm 0.0015$ (68% C.L.) [25], plotted as a function $H_i(m_X)$ for several values of $\gamma$. For $\gamma$ close to 1, the PIDM is produced shortly after reheating via freeze-in. Only for smaller values of $\gamma$ does the production during reheating become relevant. The shape of the level curves results from an interplay of the suppression of the cross section for small masses and the exponential Boltzmann suppression for large $m_X$. We also show how the contour lines are modified when including the production due to a time-dependent metric at the end of inflation (gravitational production, dotted lines), which is relevant for $m_X \simeq H_i$. The results are similar when the PIDM is a fermion, so we do not display them here.

A scalar PIDM with $m_X \ll H_i$ acquires locally a typical value $X \sim \sqrt{\langle X^2 \rangle} = H_i/2\pi$ during inflation, and the dominant energy density of the PIDM comes from the condensate generated during inflation (unless the potential of the PIDM is modified away from the pure quadratic form for large field values.). This situation leads to DM isocurvature perturbations, which are ruled out [9] (see Refs. [26–28] for related discussions). Otherwise, the PIDM inherits adiabatic perturbations from the inflaton or radiation sector.

In the limit of instantaneous reheating ($\gamma = 1$) the range of viable masses is quite large: from $10^{-10} M_p$–$10^{-2} M_p$. If reheating takes just a bit longer, then the mass range quickly shrinks significantly. For $\gamma \lesssim 10^{-3}$, corresponding to a reheating that takes $N_{rh} \gtrsim 10/(1+w)$ e-folds, only the contribution from gravitational production remains, with masses centering around $m_X \sim 10^{-6} M_p$.

It can bee seen that, indeed, there are values for which the minimal PIDM paradigm with $m_X \gtrsim M_{GUT}$ is in accordance with observations for large enough $\gamma$, corresponding to a very efficient reheating. The maximum possible mass, given that $r < 0.07$, is $0.01 M_p$. For slightly less efficient reheating this upper limit strengthens to $10^{-3} M_p (10^{-4} M_p)$ for $\gamma = 0.1(0.01)$. Therefore, the minimal PIDM setup demands that the scale of inflation is such that we expect to see tensor modes in the next round of CMB experiments. For the futuristic sensitivity of $r \sim 10^{-4}$ quoted in Ref. [24], the maximum upper bound can be improved to be $10^{-3} M_p$ for instant reheating, and $10^{-4} M_p (10^{-5} M_p)$ for $\gamma = 0.1(0.01)$. Remarkably, the entire parameter space with $m_X \gtrsim M_{GUT}$ and $\gamma \leq 1$ can be probed in the foreseeable future. Therefore, if we exclude tensors to this level in the next generation of CMB experiments, the PIDM scenario will only be viable if its mass is significantly below the natural cutoff scale. We finally reiterate that our conclusions are predicated on the standard reheating setup, with a constant equation of state and decay rate. More general scenarios could step away from these restrictions, but at the cost of introducing heavy model dependence. We checked that the results are robust when varying $w$ in the range $(-2/3, 2/3)$.

We have shown that the completely minimal model of dark matter, which is purely coupled gravitationally to all other sectors, and with mass within the ballpark of the Planck or GUT scale, is a plausible candidate. Natural values of the mass, in the sense of Occam's razor, are allowed only if the scale of inflation is high and reheating is practically instantaneous, leading to a testable prediction, that we will see primordial tensor modes in the foreseeable future. Furthermore, the required high reheating temperature allows us to discriminate a minimal PIDM from gravitationally produced superheavy dark matter, e.g., via proposed gravitational wave detectors [29]. Likewise, a determination of long reheating would allow us to conclude that there must be some additional interaction between dark matter and at least some other sector. In addition to being confirmable, this scenario is also falsifiable, in the sense that any detection of nongravitational self-interactions or interactions with the SM would immediately rule it out.

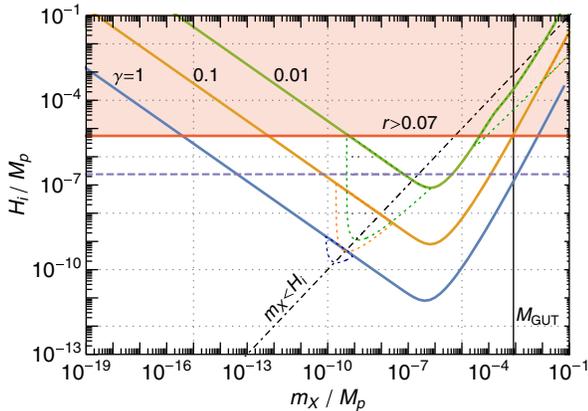

FIG. 1. The value of the Hubble rate at the start of reheating that gives the correct relic abundance, as a function of the mass of the PIDM. The blue curve is for $\gamma = 1$, the orange curve is for $\gamma = 0.1$, and the green curve is for $\gamma = 0.01$. The red region is excluded from the current bound on the tensor-to-scalar ratio, and the purple dashed line is the projected sensitivity for next generation CMB experiments, from Ref. [24]. The dotted lines show the modification when taking also gravitational production into account [12–14]. The dashed-dotted line marks $m_X = H_i$, and for a scalar PIDM the left-hand side of this line is excluded unless corrections to the PIDM potential are important during inflation. All values are given in units of $M_p$.

An interesting PIDM-like DM candidate is the Kaluza-Klein excitation of the graviton in an extra dimension with a $S_1/Z_2$ orbifold compactification. The orbifold symmetry $Z_2$ breaks Kaluza-Klein–number conservation, but leaves an invariance under parity flip of the extra dimension. The conserved Kaluza-Klein parity makes the lightest Kaluza-Klein particle stable, and if this is the first Kaluza-Klein–graviton mode, it makes an excellent PIDM candidate. Nature can have realized this in various ways. From a low energy effective point of view the simplest is to assume that the standard model lives on a three-dimensional orbifold brane, and only gravity feels the extra dimension. In string theory a similar situation could be realized if the SM lives on a stack of D3 branes. Another interesting possibility is that the PIDM is the Kaluza-Klein mode of the graviton in the 11th dimension of the Horava-Witten compactification, and in fact this is a setup where it has already been argued that there is no natural WIMP [30].

We end on a further potential signature of the PIDM: it could decay due to nonperturbative gravitational effects as in Refs. [31,32], mediated by instantons and suppressed by the Euclidean action. This leads to a flux of ultrahigh energy cosmic rays, with some overlap with the sensitivity range of the AUGER Observatory and the Telescope Array [33]. Another possibility is to detect a neutrino flux of extremely high energy, within the viable PIDM mass range [34,35]. To obtain a decay rate that yields an observable flux at present times requires a certain tuning of the instanton action, but if an exotic contribution is detected, e.g., at JEM-EUSO or at ARA [36], this could be a rare window in which it will be possible to probe the nature of quantum gravity. Finally, a peculiar feature is that, due to its high mass and its early kinetic decoupling, the PIDM has a practically zero free-streaming length; i.e., it is extremely cold, and could form substructures down to microscopic scales.

We thank Diego Blas, Daniel Figueroa, and Antonio Riotto for helpful discussions. M. S. S. is supported by a Jr. Group Leader Fellowship from the Lundbeck Foundation. The $CP^3$-Origins Center for Cosmology and Particle Physics Phenomenology is partially funded by the Danish National Research Foundation under Grant No. DNRF90.

*mathias.garny@cern.ch
†sandora@cp3.sdu.dk
‡sloth@cp3.dias.sdu.dk